\documentclass[eqsecnum,twocolumn,aps,epsf]{revtex4}
\usepackage{graphicx}
\begin{document}
\draft
\title{1/8 anomaly in the excess-oxygen-doped La$_{1.8}$Nd$_{0.2}$Cu$_{1-y}$Zn$_y$O$_{4+\delta}$}
\author{H. Mikuni, T. Adachi, S. Yairi, M. Kato and Y. Koike}
\address{Department of Applied Physics, Graduate School of Engineering, Tohoku University,\\Aoba-yama 05, Aoba-ku, Sendai 980-8579, Japan}
\author{I. Watanabe}
\address{Muon Science Laboratory, RIKEN (The Institute of Physical and Chemical Research), 2-1 Hirosawa, Wako 351-0198, Japan}
\author{K. Nagamine}
\address{Meson Science Laboratory, Institute of Materials Structure Science, High Energy Accelerator Research Organization (KEK-MSL), 1-1 Oho, Tsukuba 305-0801, Japan}

\date{\today}

\begin{abstract}

We have found the 1/8 anomaly, namely, the anomalous suppression of superconductivity at $p$ (the hole concentration per Cu) $= 1/8$ in the excess-oxygen-doped La$_{1.8}$Nd$_{0.2}$Cu$_{1-y}$Zn$_y$O$_{4+\delta}$, where the excess oxygen is doped by the electrochemical technique and the phase separation of the excess oxygen is suppressed.
The 1/8 anomaly has become marked by the 1\% substitution of Zn for Cu.
The muon-spin-relaxation measurements have revealed that the magnetic correlation develops at low temperatures below about 50 K in both Zn-free and 1\% Zn-substituted samples with $p = 1/8$.
Clear precession of muon spins suggesting the formation of a long-range magnetic order has been observed below 1 K in the 1\% Zn-substituted sample with $p = 1/8$.
These results are consistent with the stripe-pinning model.

\end{abstract}
\vspace*{2em}
\maketitle
\newpage

\section{Introduction}

The so-called 1/8 anomaly, namely, the anomalous suppression of superconductivity at $p$ (the hole concentration per Cu) $= 1/8$ in the high-$T_{\rm c}$ superconductors has attracted great interest in relation to the stripe correlations of spins and holes and also to the mechanism of the high-$T_{\rm c}$ superconductivity.
The 1/8 anomaly was discovered for the first time in La$_{2-x}$Ba$_x$CuO$_4$ (LBaCO)\cite{mooden,kumagai} which underwent a structural phase transition from the orthorhombic mid-temperature (OMT) phase (space group: Bmab) to the tetragonal low-temperature (TLT) phase (space group: P4$_2$/ncm) at a low temperature below 60 K.\cite{axe,suzuki,suzuki2}
In La$_{2-x}$Sr$_x$CuO$_4$ (LSCO) which remains in the OMT phase even at low temperatures, on the other hand, the 1/8 anomaly is not so conspicuous.
However, it was found that the 1/8 anomaly became marked in La$_{2-x}$Sr$_x$Cu$_{1-y}$Zn$_y$O$_4$ (LSCZO) by only 1\% substitution of Zn for Cu, though it remained in the OMT phase.\cite{koike}
Moreover, the 1/8 anomaly was clearly observed in La$_{1.6-x}$Nd$_{0.4}$Sr$_x$CuO$_4$ (LNSCO) which underwent the structural phase transition to the TLT phase at a low temperature below 80 K as a result of the partial substitution of Nd for La.\cite{crawford}
The origin of the 1/8 anomaly was a long-standing problem, but the discovery of a static stripe order of spins and holes in the CuO$_2$ plane from the elastic neutron-scattering experiment at $p\sim1/8$ in LNSCO by Tranquada {\it et al}. threw a new light on the 1/8 problem.\cite{tranq,tranq2}
Incommensurate magnetic peaks around ($\pi$,$\pi$) in the reciprocal lattice space similar to those observed at $p\sim1/8$ in LNSCO were observed from the elastic neutron-scattering experiment in the 3\% Zn-substituted LSCZO with $x = 0.12$ and $y = 0.03$ also.\cite{kimura}
In LSCO, moreover, such incommensurate magnetic peaks around ($\pi$,$\pi$) were also observed from the inelastic neutron-scattering experiments in a wide range of $0.06 \le x \le 0.25$ where superconductivity appeared.\cite{cheong,mason,thurston,yamada}
Accordingly, it seems that dynamical stripes, namely, dynamical stripe correlations of spins and holes exist in the CuO$_2$ plane of the La-214 system and tend to be statically stabilized at $p\sim1/8$ and are strongly pinned by the TLT structure or nonmagnetic impurities such as Zn, leading to the appearance of the static stripe order and the suppression of superconductivity at $p\sim1/8$.

In the La-214 system, it is known that holes are able to be doped into the CuO$_2$ plane not only by the partial substitution of divalent alkaline-earth elements for the trivalent La but also by incorporation of excess oxygen between the LaO-LaO planes, as shown in Fig. 1.\cite{schirber,zhou}
However, the excess-oxygen-doped La$_2$CuO$_{4+\delta}$ (LCO) is not suitable for studies of the continuous $p$ dependence of physical properties, because phase separation of the excess oxygen occurs below room temperature.\cite{jorgensen,radaelli}
The phase separation was found to be suppressed by the partial substitution of the trivalent Bi or Nd for the trivalent La on account of the random potential introduced into the LaO plane.\cite{takeda,takeda2,hiroi}
Accordingly, Kato {\it et al}. investigated the detailed $p$ dependence of the superconducting transition temperature, $T_{\rm c}$, in the excess-oxygen-doped La$_{2-x}$Bi$_x$CuO$_{4+\delta}$ (LBiCO) ($0.05\le x \le0.10$) changing $\delta$ finely, where the excess oxygen was doped by annealing under high pressures of oxygen \cite{kato,kato2} or by the KMnO$_4$ oxidation method.\cite{kato3}
They succeeded in observing a continuous increase in $T_{\rm c}$ with increasing $\delta$ in the underdoped regime, which was almost similar to that observed in LSCO.
However, the 1/8 anomaly was not observed in LBiCO, where the crystal structure remained in the tetragonal high-temperature (THT) phase (space group: I4/mmm) even at low temperatures for $p\sim1/8$.

\begin{figure}[tbp]
\begin{center}
\includegraphics[width=0.8\linewidth]{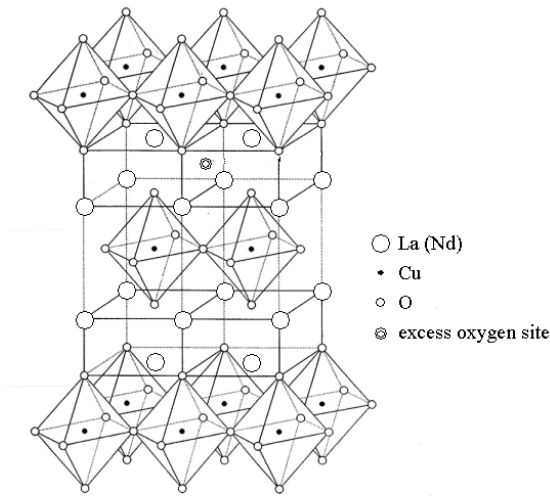}
\end{center}
\caption{Crystal structure of La$_{1.8}$Nd$_{0.2}$CuO$_{4+\delta}$.}  
\label{fig:fig1}
\end{figure}
In this paper, in order to search for the 1/8 anomaly in the excess-oxygen-doped La-214 system, we have prepared the excess-oxygen-doped La$_{1.8}$Nd$_{0.2}$Cu$_{1-y}$Zn$_y$O$_{4+\delta}$ (LNCZO), where the excess oxygen is doped by the electrochemical technique and the phase separation of the excess oxygen is suppressed by the 10\% substitution of Nd for La.
Here, Zn atoms are introduced so as to pin the possible dynamical stripes.
We have investigated the thermoelectric power, electrical resistivity and $T_{\rm c}$ of the excess-oxygen-doped LNCZO with $y$ = 0 and 0.01, changing $\delta$ finely.
Muon-spin-relaxation ($\mu$SR) measurements have been performed to study the Cu-spin state.
Powder x-ray diffraction measurements have also been carried out in a wide temperature-range between 10 K and room temperature to study the crystal structure.

\section{Experimental}

Sintered samples of the Zn-free ($y$ = 0) and 1\% Zn-substituted ($y$ = 0.01) La$_{1.8}$Nd$_{0.2}$Cu$_{1-y}$Zn$_y$O$_4$ were prepared by the solid-state reaction method from appropriated powders of La$_2$O$_3$, Nd$_2$O$_3$, CuO and ZnO.
The powders were mixed and prefired in air at 900$^{\circ}$C for 12 h.
After pulverization, the prefired materials were mixed, pelletized and sintered in air at 1050$^{\circ}$C for 24 h, followed by furnace cooling.
This sintering process was carried out once again to obtain homogeneous samples.
Before the incorporation of excess oxygen, the sintered samples were annealed in flowing Ar-gas at 700$^{\circ}$C for 48 h to remove excess oxygen incorporated during the sintering process.
These samples were characterized by powder x-ray diffraction at room temperature to be of the single phase and confirmed by iodometric titration to contain no excess-oxygen within the experimental accuracy.

The excess-oxygen doping was made by the electrochemical oxidation method,\cite{greinier} using a potentio-galvanostat (Hokuto Denko Co., Model HABF-501).
A three-electrode cell was set up as La$_{1.8}$Nd$_{0.2}$Cu$_{1-y}$Zn$_y$O$_4$$\mid$aqueous solution of KOH$\mid$Pt, where the working electrode was a pellet of La$_{1.8}$Nd$_{0.2}$Cu$_{1-y}$Zn$_y$O$_4$ with the dimensions of $\sim10$ mm in diameter and $\sim1$ mm in thickness and the counter electrode was a Pt wire.
As an electrolyte, an aqueous solution of KOH (1.0 mol/l) was used at 60$^{\circ}$C.
A Hg/Hg$_2$Cl$_2$ electrode was used as a reference electrode connected to the aqueous solution of KOH via the salt bridge.
The excess-oxygen doping was carried out at a constant voltage of $\sim300$ mV.
After the excess-oxygen doping, the pellet was washed with anhydrous ethanol and dried in air at 100$^{\circ}$C for 24 h.
The amount of the excess oxygen doped into the pellet was estimated by the Faraday's law of electrolysis.
That is, it was calculated from the amount of the electrical charge which flowed in the pellet, based upon the following chemical reaction.

La$_{1.8}$Nd$_{0.2}$Cu$_{1-y}$Zn$_y$O$_4$ + 2$\delta$OH$^-$ $\rightarrow$ La$_{1.8}$Nd$_{0.2}$Cu$_{1-y}$Zn$_y$O$_{4+\delta}$ + $\delta$H$_2$O + 2$\delta$e$^-$     (1)

\noindent
Values of the excess oxygen estimated thus were confirmed to be almost the same within the experimental accuracy as those estimated by iodometric titration.

Powder x-ray diffraction measurements were made using a conventional diffractometer with a curved graphite monochromator for CuK$_{\alpha}$ radiation.
The data were analyzed after CuK$_{\alpha2}$ stripping.
Thermoelectric power measurements were carried out by the dc method with a temperature gradient of $\sim0.5$ K across the sample.
Electrical resistivity measurements were carried out by the dc four-probe method. 
$\mu$SR measurements were performed in zero field at low temperatures down to 0.3 K at the RIKEN-RAL Muon Facility at the Rutherford-Appleton Laboratory in the UK.
A pulsed positive surface-muon beam with a momentum of 27 MeV/c was used.
The asymmetry parameter, $A(t)$, was defined as $A(t) = [F(t)-\alpha B(t)]/[F(t)+\alpha B(t)]$, where $F(t)$ and $B(t)$ were total muon events counted by the forward and backward counters at a time $t$, respectively.
The $\alpha$ is a calibration factor reflecting the relative counting efficiencies of the forward and backward counters.

\section{Results}
\subsection{Crystal structure}
\begin{figure}[tbp]
\begin{center}
\includegraphics[width=0.8\linewidth]{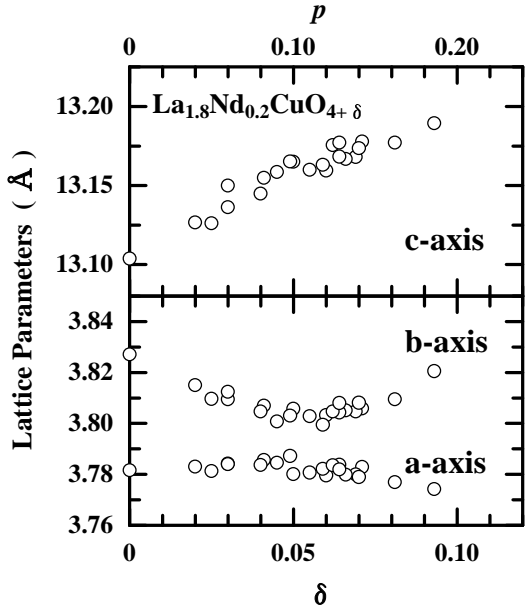}
\end{center}
\caption{Lattice parameters $a$, $b$ and $c$ at room temperature as a function of excess-oxygen content $\delta$ for La$_{1.8}$Nd$_{0.2}$CuO$_{4+\delta}$. The hole concentration per Cu, $p$, is also described on the upper transverse axis, using the equation $p = 2\delta$.}  
\label{fig:fig2} 
\end{figure}
Figure 2 shows the variation of the lattice parameters with $\delta$ at room temperature obtained from the powder x-ray diffraction measurements.
The crystal structure is orthorhombic for all samples with $0 \le \delta < 0.10$.
The lattice parameter $c$ increases monotonically with increasing $\delta$, suggesting that the excess oxygen is continuously inserted between the LaO-LaO planes.
As for the homogeneity of the excess oxygen, it is naively guessed that the excess-oxygen content is larger in the surface area of the pellet in contact with the aqueous solution of KOH than in the internal part.
However, the $c$ value does not change between the surface area and the internal part, so that the excess oxygen is concluded to be homogeneously doped into the pellet within the accuracy of the powder x-ray diffraction measurements.
It is found that the orthorhombicity decreases with increasing $\delta$ for $\delta < 0.05$, while it increases for $\delta > 0.05$.
The space group of the orthorhombic phase for $\delta < 0.05$ is guessed to be Bmab, as in the case of the lightly excess-oxygen-doped regime of LCO\cite{jorgensen,radaelli} and LBiCO.\cite{kato,kato3}
For $\delta > 0.05$, on the other hand, the orthorhombic phase with the space group Fmmm may appear, as in the case of the heavily excess-oxygen-doped regime of LCO\cite{jorgensen,radaelli} and LBiCO.\cite{hiroi,kato,kato3}.

\begin{figure}[tbp]
\begin{center}
\includegraphics[width=1.0\linewidth]{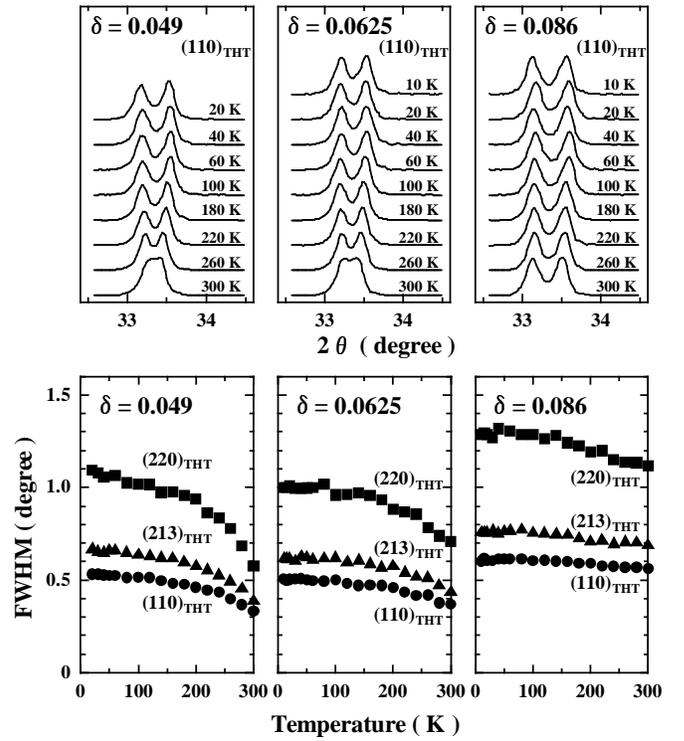}
\end{center}
\caption{(Upper) Powder x-ray diffraction profiles of the (110)$_{THT}$ reflection due to CuK$_{\alpha}$ radiation at various temperatures for $\delta$ = 0.049, 0.0625 and 0.086 in La$_{1.8}$Nd$_{0.2}$CuO$_{4+\delta}$. (Lower) Temperature dependence of the FWHM of the (110)$_{THT}$, (213)$_{THT}$ and (220)$_{THT}$ peaks due to CuK$_{\alpha}$ radiation for $\delta$ = 0.049, 0.0625 and 0.086 in La$_{1.8}$Nd$_{0.2}$CuO$_{4+\delta}$.}  
\label{fig:fig3} 
\end{figure}
Upper panels of Fig. 3 display the powder x-ray diffraction profiles of the (110)$_{THT}$ reflection at various temperatures for the typical samples with $\delta = 0.049, 0.0625$ and 0.086.
The suffix THT indicates an index in the THT phase.
All the (110)$_{THT}$ peaks are found to clearly split into two, meaning that the crystal structure is orthorhombic.
The orthorhombicity is found to increase with decreasing temperature for these samples.
This is clearly seen in the lower panels of Fig. 3, where the temperature dependence of the full width at half-maximum (FWHM) of the (110)$_{THT}$, (213)$_{THT}$ and (220)$_{THT}$ peaks is shown.
The increase of the orthorhombicity with decreasing temperature is most marked in $\delta$ = 0.049 among these samples and is comparable to that generally observed in the OMT phase of the La-214 system.
This is consistent with the above guess that the space group for $\delta < 0.05$ is Bmab.
Moreover, what is significant is that neither symptom of the phase transition to the TLT phase nor the phase separation of the excess oxygen are observed at low temperatures down to 10 K for these samples.

\subsection{Thermoelectric power and electrical resistivity}
\begin{figure}[tbp]
\begin{center}
\includegraphics[width=0.9\linewidth]{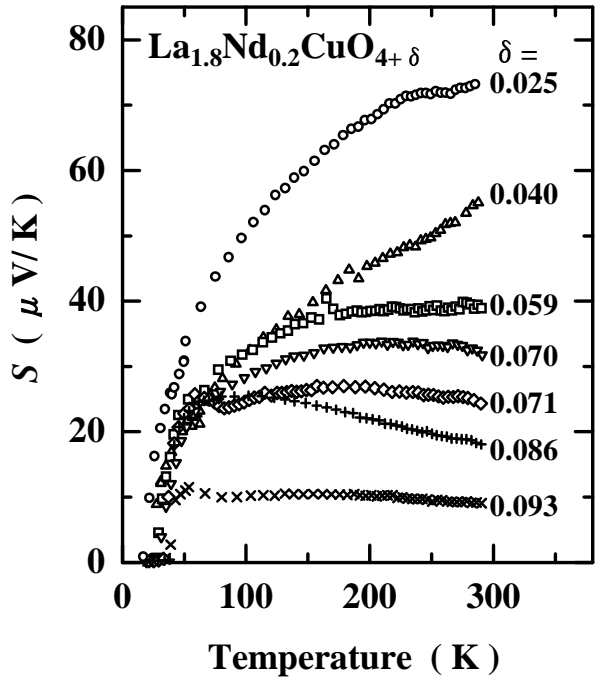}
\end{center}
\caption{Temperature dependence of the thermoelectric power $S$ for various $\delta$ values in La$_{1.8}$Nd$_{0.2}$CuO$_{4+\delta}$.}  
\label{fig:fig4} 
\end{figure}
Figure 4 displays the temperature dependence of the thermoelectric power $S$ for various $\delta$ values.
The value of $S$ decreases with increasing $\delta$.
Taking $p = 2\delta$, this is a typical behavior observed in the hole-doped high-$T_{\rm c}$ cuprates.\cite{obertelli}
In fact, the $\delta$ dependence of the value of $S$ at 290 K, $S_{290K}$, is similar to that of LSCO,\cite{adachi} as shown in Fig. 5.
This suggests that the CuO$_2$ plane is continuously supplied with holes with increasing $\delta$.
\begin{figure}[tbp]
\begin{center}
\includegraphics[width=0.9\linewidth]{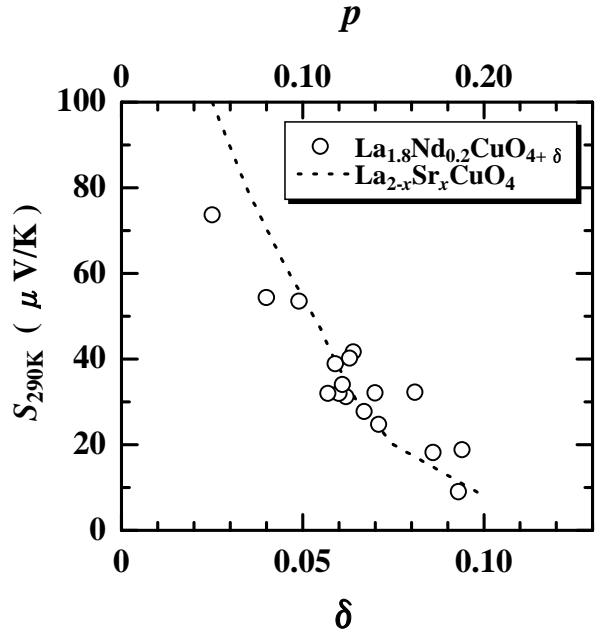}
\end{center}
\caption{Dependence on $\delta$ of $S_{290K}$ for La$_{1.8}$Nd$_{0.2}$CuO$_{4+\delta}$.  The hole concentration per Cu, $p$, is also described on the upper transverse axis, using the equation $p = 2\delta$. The dashed line indicates the data of La$_{2-x}$Sr$_x$CuO$_4$ for reference.\cite{adachi}}  
\label{fig:fig5} 
\end{figure}

\begin{figure}[tbp]
\begin{center}
\includegraphics[width=0.9\linewidth]{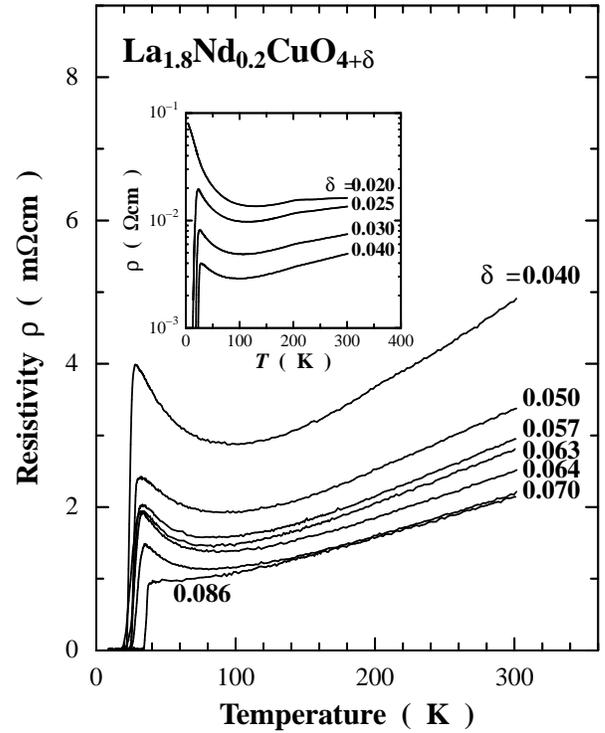}
\end{center}
\caption{Temperature dependence of the electrical resistivity $\rho$ for various $\delta$ values in La$_{1.8}$Nd$_{0.2}$CuO$_{4+\delta}$. The inset shows the temperature dependence of $\rho$ in the logarithmic scale for the lightly doped samples.}  
\label{fig:fig6} 
\end{figure}
Figure 6 displays the temperature dependence of the electrical resistivity $\rho$ for various $\delta$ values.
The value of $\rho$ decreases with increasing $\delta$, namely, with increasing $p$.
A metallic behavior (d$\rho$/d$T > 0$) is observed in the high-temperature region, while an upturn behavior of $\rho$ (d$\rho$/d$T < 0$) is observed at low temperatures below about 100 K for $\delta \le 0.07$.
The superconducting transition appears for $\delta \ge 0.025$.
The $\delta$ dependence of $T_{\rm c}$, defined as the midpoint of the superconducting transition curve in the $\rho$ vs $T$ plot, is shown in the lower panel of Fig. 7.
The $\delta$ dependences of $\rho$ and $T_{\rm c}$ are different from those of LCO where the phase separation of the excess oxygen occurs.\cite{mikuni}
Taking $p = 2\delta$, the $p$ dependences of $\rho$ and $T_{\rm c}$ are similar to those of LSCO, respectively.
These results also suggest that the phase separation of the excess oxygen is suppressed and that the doped excess-oxygen supplies the CuO$_2$ plane with holes homogeneously.
What is remarkable is that a small dip of $T_{\rm c}$ is observed in the $\delta$ dependence of $T_{\rm c}$ around $\delta$ = 0.0625 ($p$ = 1/8), which is very similar to that observed in LSCO around $p$ = 1/8.\cite{koike}
It is noted that the superconductivity at $\delta = 0.0625$ is confirmed to be of the bulk from the magnetic susceptibility measurement.~\cite{mikuni}
The upturn behavior of $\rho$ is generally observed at low temperatures in the underdoped regime of the high-$T_{\rm c}$ cuprates, which is understood to be due to localization of carriers.
The degree of the localization may be estimated by the value of $(\rho_{max}-\rho_{min})/\rho_{min}$, where $\rho_{min}$ and $\rho_{max}$ are defined as the minimum value of $\rho$ around 100 K and the maximum value of $\rho$ just above $T_{\rm c}$, respectively.
As shown in the upper panel of Fig. 7, $(\rho_{max}-\rho_{min})/\rho_{min}$ roughly decreases with increasing $\delta$, namely, with increasing $p$.
However, a small hump is observed around $\delta$ = 0.0625, which is similar to that observed around $p$ = 1/8 in LBaCO.\cite{koike2}
Therefore, the small dip of $T_{\rm c}$ around $\delta$ = 0.0625 is regarded as the very 1/8 anomaly.
\begin{figure}[tbp]
\begin{center}
\includegraphics[width=0.9\linewidth]{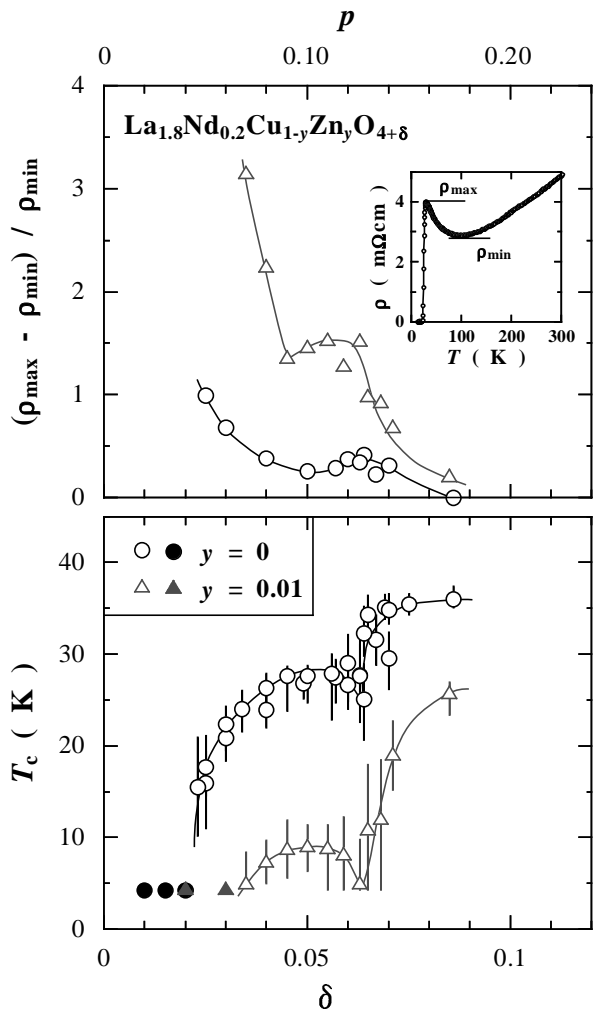}
\end{center}
\caption{Dependences on $\delta$ of $T_{\rm c}$ and $(\rho_{max}-\rho_{min})/\rho_{min}$ for the Zn-free ($y$ = 0) and 1\% Zn-substituted ($y$ = 0.01) La$_{1.8}$Nd$_{0.2}$Cu$_{1-y}$Zn$_y$O$_{4+\delta}$. The $T_{\rm c}$ is defined as the midpoint of the superconducting transition curve in the $\rho$ vs $T$ plot. Bars indicate temperatures where $\rho$ drops to 90\% and 10\% of the normal-state value of $\rho$. Closed symbols indicate samples whose $T_{\rm c}$'s are not above 4.2 K. The $\rho_{min}$ and $\rho_{max}$ are defined as the minimum value of $\rho$ around 100 K and the maximum value of $\rho$ just above $T_{\rm c}$, respectively, as shown in the inset. The hole concentration per Cu, $p$, is also described on the upper transverse axis, using the equation $p = 2\delta$. Lines are guides to the eye.}  
\label{fig:fig7} 
\end{figure}

The same measurements were carried out in the 1\% Zn-substituted samples also.
As shown in the upper panel of Fig. 7, $(\rho_{max}-\rho_{min})/\rho_{min}$ increases as a result of the 1\% Zn-substitution for each $\delta$.
It exhibits a hump around $\delta$ = 0.0625 in the Zn-substituted samples as well as in the Zn-free ones.
On the other hand, $T_{\rm c}$ decreases as a result of the 1\% Zn-substitution for each $\delta$.
The dip of $T_{\rm c}$ around $\delta$ = 0.0625 becomes marked in the Zn-substituted samples, as in the case of LSCZO.\cite{koike}
It is noted from the magnetic susceptibility measurement that the superconductivity in the Zn-subsutituted sample with $\delta = 0.0625$ is not of the bulk, compared with the bulk superconductivity in the Zn-free sample with $\delta = 0.0625$.~\cite{mikuni}
Accordingly, it appears that the 1\% Zn-substitution enhances the 1/8 anomaly.

\subsection{Muon spin relaxation}
\begin{figure}[bp]
\begin{center}
\includegraphics[width=0.9\linewidth]{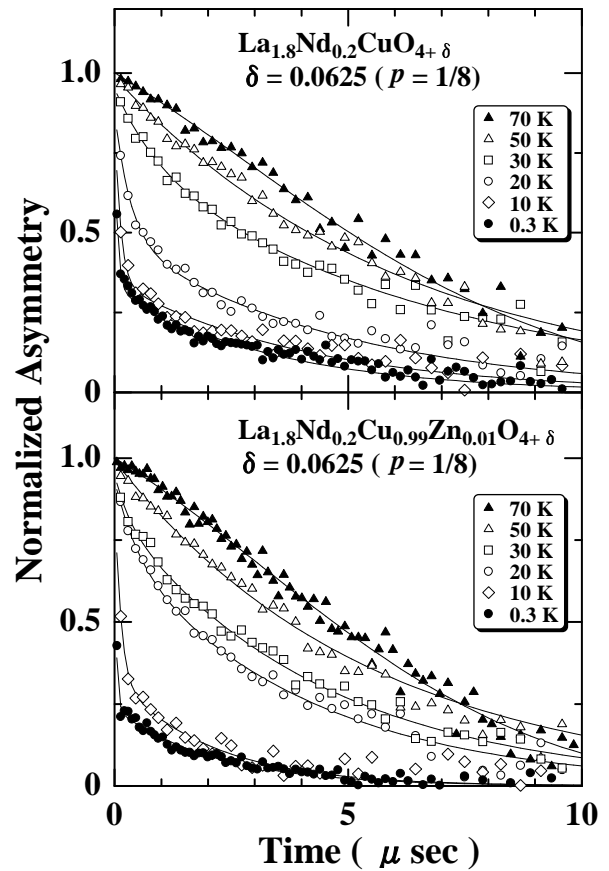}
\end{center}
\caption{Zero-field $\mu$SR time spectra at various temperatures for the Zn-free ($y$ = 0) and 1\% Zn-substituted ($y$ = 0.01) La$_{1.8}$Nd$_{0.2}$Cu$_{1-y}$Zn$_y$O$_{4+\delta}$ with $\delta$ = 0.0625 ($p$ = 1/8). Solid lines indicate the best-fit results using $A(t) = A_0$e$^{-\lambda_0t}G_z(\Delta,t) + A_1$e$^{-\lambda_1t}$.}
\label{fig:fig8} 
\end{figure}
Figure 8 displays the $\mu$SR time spectra, namely, the time evolution of $A(t)$ in zero field for the Zn-free and 1\% Zn-substituted samples with $\delta$ = 0.0625 ($p$ = 1/8).
In both Zn-free and Zn-substituted samples, the time spectrum shows a Gaussian-type depolarization behavior at high temperatures above $\sim60$ K, indicating that the Cu spins are fluctuating so fast as not to affect the muon spins.
In the both samples, the depolarization behavior deviates from the Gaussian-type at low temperatures below $\sim60$ K and a fast depolarizing component appears, meaning that the Cu-spin fluctuations exhibit a slowing-down behavior on account of the development of the magnetic correlation at low temperatures.
Even at the lowest temperature of 0.3 K, the dynamical long-time relaxation behavior is observed, which may be due to the Nd moments.
In the Zn-substituted sample, muon-spin precession is observed at very low temperatures below 1 K, as clearly seen in Fig. 9, indicating the appearance of a long-range ordered state of Cu spins.
\begin{figure}[tbp]
\begin{center}
\includegraphics[width=0.9\linewidth]{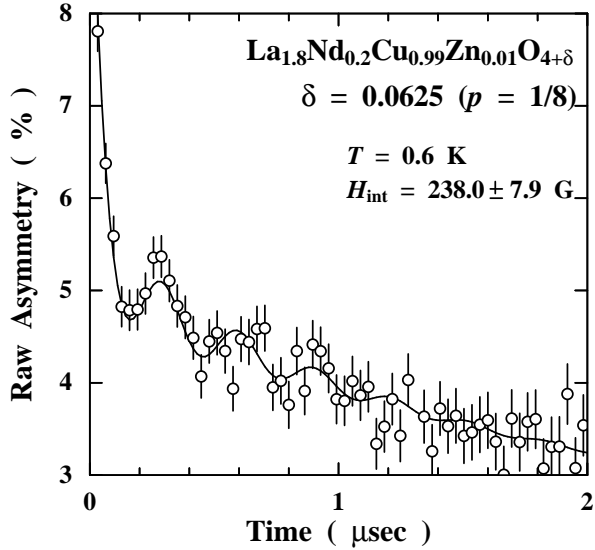}
\end{center}
\caption{Zero-field $\mu$SR time spectrum of the 1\% Zn-substituted ($y$ = 0.01) La$_{1.8}$Nd$_{0.2}$Cu$_{1-y}$Zn$_y$O$_{4+\delta}$ with $\delta$ = 0.0625 ($p$ = 1/8) at 0.6 K. The solid line indicates the best-fit result using $A(t) = A_0$e$^{-\lambda_0t} + A_1$e$^{-\lambda_1t} + A_2$e$^{-\lambda_2t}$cos$(\omega t + \phi)$. The internal field at the muon site, $H_{\rm int}$, is estimated as $238.0 \pm 7.9$ gauss.}
\label{fig:fig9} 
\end{figure}

The two-component function $A(t) = A_0$e$^{-\lambda_0t}G_z(\Delta,t) + A_1$e$^{-\lambda_1t}$ is used conventionally in order to analyze the time spectra following our previous works.~\cite{watanabe5,watanabe4}
The first term indicates the slowly depolarizing component of muon spins due to the nuclear dipole fields.
Here, $A_0$ and $\lambda_0$ are the initial asymmetry and the depolarization rate of the slowly depolarizing component, respectively, and $G_z(\Delta,t)$ is the Kubo-Toyabe function representing the effect of the nuclear dipole fields distributed at the muon site with a distribution width of $\Delta$.\cite{uemura}
The second term indicates the fast depolarizing component of muon spins, where $A_1$ and $\lambda_1$ are the initial asymmetry and the depolarization rate of the fast depolarizing component, respectively.
The Kubo-Toyabe function is no longer valid when the internal field due to electronic moments becomes dominant at the muon site rather than the weak field due to nuclear dipoles near the magnetic transition temperature.
In this case, the value of $\Delta$ becomes zero effectively, resulting in $G_z(\Delta,t)=1$.
The best-fit results are shown by solid lines in Fig. 8.
The temperature dependence of $A_0$ obtained from the best fitting is displayed in Fig. 10.
In both Zn-free and 1\% Zn-substituted samples, $A_0$ markedly decreases with decreasing temperature below $\sim50$ K and is constant at low temperatures below $\sim5$ K.
This is a typical change of $A_0$ in the case where a static magnetically ordered state of Cu spins appears in the La-214 system.\cite{watanabe,watanabe2}
In the both samples, the saturated value of $A_0$ at low temperatures below $\sim5$ K is about one third of that at high temperatures, meaning that all of the muon spins in the samples feel spontaneous internal field.
That is to say, all of the Cu spins are in the magnetically ordered state.\cite{watanabe2,watanabe3}
The magnetic transition temperature $T_{\rm N}$, defined as the midpoint of the transition in the $A_0$ vs $T$ plot, is 25 K and 17 K for the Zn-free and 1\% Zn-substituted samples, respectively.
\begin{figure}[tbp]
\begin{center}
\includegraphics[width=0.78\linewidth]{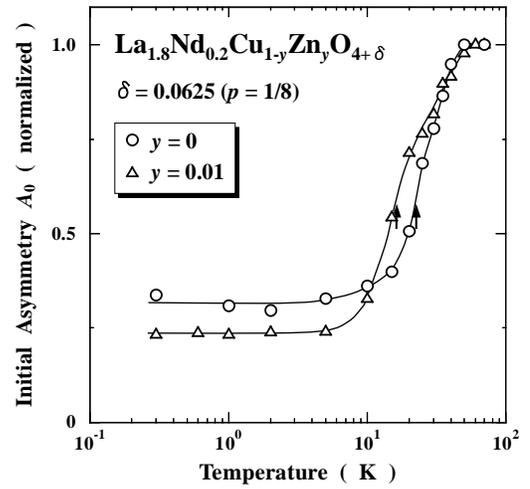}
\end{center}
\caption{Temperature dependence of the initial asymmetry of the slowly depolarizing component $A_0$ for the Zn-free ($y$ = 0) and 1\% Zn-substituted ($y$ = 0.01) La$_{1.8}$Nd$_{0.2}$Cu$_{1-y}$Zn$_y$O$_{4+\delta}$ with $\delta$ = 0.0625 ($p$ = 1/8). Arrows indicate the magnetic transition temperature, $T_{\rm N}$, defined as the midpoint of the transition in the $A_0$ vs $T$ plot.}
\label{fig:fig10} 
\end{figure}

The time spectrum exhibiting the muon-spin precession for the 1\% Zn-substituted sample shown in Fig. 9 is fitted using the three-component function $A(t) = A_0$e$^{-\lambda_0t} + A_1$e$^{-\lambda_1t} + A_2$e$^{-\lambda_2t}$cos$(\omega t + \phi)$, where $A_2$ and $\lambda_2$ are the amplitude and the damping rate of the muon-spin precession component, respectively.
The $\omega$ and $\phi$ are the frequency and the phase of the muon-spin precession at $t$ = 0, respectively.
From the best fitting, the internal field at the muon site, $H_{\rm int}$, is estimated from the relation $H_{\rm int} = \omega / \gamma_{\mu}$ to be $238.0 \pm 7.9$ gauss.
Here, $\gamma_{\mu}$ is the gyromagnetic ratio of the muon spin ($\gamma_{\mu} / 2 \pi = 13.55$ MHz/kOe).
This is much smaller than that in La$_2$CuO$_4$ (420 gauss)\cite{uemura2} and almost the same as that in the 1\% Zn-substituted LSCZO with $p \sim 1/8$ ($\sim 220$ gauss).\cite{watanabe3,watanabe4}

\section{Discussion}

We have found the 1/8 anomaly and also a magnetic transition at $\delta$ = 0.0625 ($p$ = 1/8) in both Zn-free and 1\% Zn-substituted LNCZO.
Moreover, the internal field at low temperatures in the 1\% Zn-substituted LNCZO with $\delta$ = 0.0625 is almost the same as that in the 1\% Zn-substituted LSCZO with $p \sim 1/8$, suggesting that the spin structures are similar to each other.
Accordingly, it is very likely that the static stripe order is formed at $\delta$ = 0.0625 in both Zn-free and 1\% Zn-substituted LNCZO as well as in LSCZO.
The 1/8 anomaly has been found to become marked by the 1\% substitution of Zn for Cu in LNCZO, as in the case of LSCZO.
Furthermore, the muon-spin precession is clearly observed in the Zn-substituted sample rather than in the Zn-free one, indicating that the magnetic order in the Zn-substituted sample is more long-ranged than that in the Zn-free one.
These results are consistent with the stripe-pinning model, where the dynamical stripes are regarded as existing in the CuO$_2$ plane and tending to be statically stabilized at $p\sim1/8$ and being effectively pinned by a small amount of Zn, leading to the appearance of the static stripe order and the suppression of superconductivity.\cite{adachi,koike3,koike4}
As for the difference in $T_{\rm N}$ between the Zn-free and 1\% Zn-substituted samples, the effect of the spin dilution due to the substitution of the nonmagnetic Zn for the magnetic Cu may emerge at relatively high temperatures above $\sim10$ K in the Zn-substituted sample, leading to the decrease in $T_{\rm N}$.
A similar effect by the Zn substitution has also been observed around $p = 1/8$ in the Zn-substituted LSCZO.\cite{adachi2}
Once the Cu-spin fluctuations slow down at relatively low temperatures below $\sim10$ K, on the other hand, Zn may operate to pin the dynamical stripes strongly, leading to the formation of a more long-ranged order in the Zn-substituted sample than in the Zn-free one.

Although the 1/8 anomaly has been found in the Zn-free LNCZO, no 1/8 anomaly has been observed in the excess-oxygen-doped LBiCO, as mentioned in Section I.\cite{kato,kato2,kato3}
According to the stripe-pinning model, the reason may be as follows.
The buckling of the CuO$_2$ plane in the TLT structure is known to be effective for the pinning of the dynamical stripes, because the rotation axis of the buckling is parallel to the stripes.
On the other hand, the flatness of the CuO$_2$ plane is considered to be ineffective for the pinning.
LBiCO around $p\sim1/8$ has the THT structure where the CuO$_2$ plane is flat.
Therefore, the dynamical stripes are hard to be statically stabilized, leading to no 1/8 anomaly in LBiCO.
Actually, in thin films of LBaCO where the CuO$_2$ plane is so compressed due to the lattice mismatch between the film and substrate as to be probably flat, no 1/8 anomaly has been observed, though it is clearly observed in bulk samples of LBaCO with the TLT structure.\cite{sato}
Accordingly, the 1/8 anomaly observed in the Zn-free LNCZO and LSCO with the orthorhombic structure may be explained as being due to the buckling of the CuO$_2$ plane, though the rotation axis of the buckling is different from that in the TLT structure.

Finally, we comment on the staging structure of the excess oxygen, which is known to be formed in the excess-oxygen-doped LCO.\cite{wells,wells2,blakeslee}
In the present LNCZO, however, no staging structure has been detected within the accuracy of the powder x-ray diffraction measurements.
Neutron scattering experiments are necessary to check the staging structure in detail.
However, even if any staging structure is formed, holes must be homogeneously distributed in the CuO$_2$ plane, because the $p$ dependences of $S$, $\rho$ and $T_{\rm c}$ are similar to those of LSCO.

\section{Summary}

We have found the 1/8 anomaly in the excess-oxygen-doped LNCZO, where the excess oxygen is doped by the electrochemical technique and the phase separation of the excess oxygen is suppressed.
The 1/8 anomaly has become marked by the 1\% substitution of Zn for Cu.
It has been found from the $\mu$SR measurements that the magnetic correlation develops at low temperatures below about 50 K in both Zn-free and 1\% Zn-substituted samples with $p = 1/8$.
The magnetic transition temperature is estimated as 25 K and 17 K for the Zn-free and Zn-substituted samples, respectively.
Clear precession of muon spins has been observed below 1 K in the 1\% Zn-substituted sample with $p = 1/8$ rather than in the Zn-free sample, indicating that the magnetic order in the Zn-substituted sample is more long-ranged than that in the Zn-free sample.
These results are consistent with the stripe-pinning model where the dynamical stripes are regarded as existing in the CuO$_2$ plane and tending to be statically stabilized at $p\sim1/8$ and effectively pinned by a small amount of nonmagnetic impurities such as Zn, leading to the appearance of the static stripe order and the suppression of superconductivity at $p\sim1/8$.
Compared with the result in the excess-oxygen-doped LBiCO, it appears that the buckling of the CuO$_2$ plane is indispensable to the appearance of the 1/8 anomaly.

\section*{Acknowledgments}

This work was supported by a Grant-in-Aid for Scientific Research from the Ministry of Education, Culture, Sports, Science and Technology, Japan, and also by CREST of Japan Science and Technology Corporation.


\begin{references}

\bibitem{mooden}
A. R. Moodenbaugh, Youwen Xu, M. Suenaga, T. J. Folkerts, and R. N. Shelton,
 Phys. Rev. B {\bf 38}, 4596 (1988).

\bibitem{kumagai}
K. Kumagai, Y. Nakamura, I. Watanabe, Y. Nakamichi, and H. Nakajima,
 J. Magn. Magn. Mater. {\bf 76\&77}, 601 (1988).

\bibitem{axe}
J. D. Axe, A. H. Moudden, D. Hohlwein, D. E. Cox, K. M. Mohanty, A. R. Moodenbaugh, and Youwen Xu,
 Phys. Rev. Lett. {\bf 62}, 2751 (1989).

\bibitem{suzuki}
T. Suzuki and T. Fujita,
 J. Phys. Soc. Jpn. {\bf 58}, 1883 (1989).

\bibitem{suzuki2}
T. Suzuki and T. Fujita,
 Physica C {\bf 159}, 111 (1989).

\bibitem{koike}
Y. Koike, A. Kobayashi, T. Kawaguchi, M. Kato, T. Noji, Y. Ono, T. Hikita, and Y. Saito,
 Solid State Commun. {\bf 82}, 889 (1992).

\bibitem{crawford}
M. K. Crawford, R. L. Harlow, E. M. McCarron, W. E. Farneth, J. D. Axe, H. Chou, and Q. Huang,
 Phys. Rev. B {\bf 44}, 7749 (1991).

\bibitem{tranq}
J. M. Tranquada, B. J. Sternlieb, J. D. Axe, Y. Nakamura, and S. Uchida,
 Nature (London) {\bf 375}, 561 (1995).

\bibitem{tranq2}
J. M. Tranquada, J. D. Axe, N. Ichikawa, Y. Nakamura, S. Uchida, and B. Nachumi,
 Phys. Rev. B {\bf 54}, 7489 (1996).

\bibitem{kimura}
H. Kimura, K. Hirota, H. Matsushita, K. Yamada, Y. Endoh, S. -H. Lee, C. F. Majkrzak, R. Erwin, G. Shirane, M. Greven, Y. S. Lee, M. A. Kastner, and R. J. Birgeneau,
 Phys. Rev. B {\bf 59}, 6517 (1999).

\bibitem{cheong}
S. -W. Cheong, G. Aeppli, T. E. Mason, H. Mook, S. M. Hayden, P. C. Canfield, Z. Fisk, K. N. Clausen, and J. L. Martinez,
 Phys. Rev. Lett. {\bf 67}, 1791 (1991).

\bibitem{mason}
T. E. Mason, G. Aeppli, and H. A. Mook,
 Phys. Rev. Lett. {\bf 68}, 1414 (1992).

\bibitem{thurston}
T. R. Thurston, P. M. Gehring, G. Shirane, R. J. Birgeneau, M. A. Kastner, Y. Endoh, M. Matsuda, K. Yamada, H. Kojima, and I. Tanaka,
 Phys. Rev. B {\bf 46}, 9128 (1992).

\bibitem{yamada}
K. Yamada, C. H. Lee, K. Kurahashi, J. Wada, S. Wakimoto, S. Ueki, H. Kimura, Y. Endoh, S. Hosoya, G. Shirane, R. J. Birgeneau, M. Greven, M. A. Kastner, and Y. J. Kim,
 Phys. Rev. B {\bf 57}, 6165 (1998).

\bibitem{schirber}
J. E. Schirber, B. Morosin, R. M. Merrill, P. F. Hlava, E. L. Venturini, J. F. Kwak, P. J. Nigrey, R. J. Baughman, and D. S. Ginley,
 Physica C {\bf 152}, 121 (1988).

\bibitem{zhou}
J. Zhou, S. Sinha, and J. B. Goodenough,
 Phys. Rev. B {\bf 39}, 12331 (1989).

\bibitem{jorgensen}
J. D. Jorgensen, P. Lightfoot, S. Pei, B. Dabrowski, D. R. Richards, and D. G. Hinks,
 in: Advances in Superconductivity III (Springer, Berlin, 1990) p. 337.

\bibitem{radaelli}
P. G. Radaelli, J. D. Jorgensen, R. Kleb, B. A. Hunter, F. C. Chou, and D. C. Johnston,
 Phys. Rev. B {\bf 49}, 6239 (1994).

\bibitem{takeda}
Y. Takeda, K. Yoshikawa, N. Imanishi, O. Yamamoto, and M. Takano,
 J. Solid State Chem. {\bf 92}, 241 (1991).

\bibitem{takeda2}
Y. Takeda, A. Sato, K. Yoshikawa, N. Imanishi, O. Yamamoto, M. Takano, Z. Hiroi, and Y. Bando,
 Physica C {\bf 185-189}, 603 (1991).

\bibitem{hiroi}
Z. Hiroi, M. Takano, Y. Bando, A. Sato, and Y. Takeda,
 Phys. Rev. B {\bf 46}, 14857 (1992).

\bibitem{kato}
M. Kato, T. Aoki, T. Noji, Y. Ono, Y. Koike, T. Hikita, and Y. Saito,
 Physica C {\bf 217}, 189 (1993).

\bibitem{kato2}
M. Kato, T. Aoki, H. Chizawa, Y. Koike, Y. Ono, T. Noji, T. Hikita, and Y. Saito,
 Physica C {\bf 235-240}, 337 (1994).

\bibitem{kato3}
M. Kato, H. Chizawa, Y. Ono, and Y. Koike,
 Physica C {\bf 256}, 253 (1996).

\bibitem{greinier}
J. C. Greinier, A. Wattiaux, N. Laugueyte, J. C. Park, E. Marguestaut, J. Etourneau, and M. Pouchard,
 Physica C {\bf 173}, 139 (1991).

\bibitem{obertelli}
S. D. Obertelli, J. R. Cooper, and J. L. Tallon,
 Phys. Rev.B {\bf 46}, 14928 (1992).

\bibitem{adachi}
T. Adachi, T. Noji, H. Sato, Y. Koike, T. Nishizaki, and N. Kobayashi,
 J. Low Temp. Phys. {\bf 117}, 1151 (1999).

\bibitem{mikuni}
H. Mikuni, T. Adachi, and Y. Koike,
 unpublished.

\bibitem{koike2}
Y. Koike, T. Kawaguchi, N. Watanabe, T. Noji and Y. Saito,
 Solid State Commun. {\bf 79}, 155 (1991).

\bibitem{watanabe5}
I. Watanabe, M. Aoyama, M. Akoshima, T. Kawamata, T. Adachi, Y. Koike, S. Ohira, W. Higemoto, and K. Nagamine,
 Phys. Rev. B {\bf 62}, R11985 (2000).

\bibitem{watanabe4}
I. Watanabe, T. Adachi, K. Takahashi, S. Yairi, Y. Koike, and K. Nagamine,
 Phys. Rev. B {\bf 65}, 180516 (2002).

\bibitem{uemura}
Y. J. Uemura, T. Yamazaki, D. R. Harshman, M. Senba, and E. J. Ansaldo,
 Phys. Rev. B {\bf 31}, 546 (1985).

\bibitem{watanabe}
I. Watanabe, K. Kawano, K. Kumagai, K. Nishiyama, and K. Nagamine,
 J. Phys. Soc. Jpn. {\bf 61}, 3058 (1992).

\bibitem{watanabe2}
I. Watanabe, K. Nishiyama, K. Nagamine, K. Kawano, and K. Kumagai,
 Hyperfine Interact. {\bf 86}, 603 (1994).

\bibitem{watanabe3}
I. Watanabe, T. Adachi, K. Takahashi, S. Yairi, Y. Koike, and K. Nagamine,
 J. Phys. Chem. Solids {\bf 63}, 1093 (2002).

\bibitem{uemura2}
Y. J. Uemura, W. J. Kossler, X. H. Yu, J. R. Kempton, H. E. Schone, D. Opie, C. E. Stronach, D. C. Johnston, M. S. Alvarez, and D. P. Goshorn,
 Phys. Rev. Lett. {\bf 59}, 1045 (1987).

\bibitem{koike3}
Y. Koike, S. Takeuchi, H. Sato, Y. Hama, M. Kato, Y.Ono, and S. Katano,
 J. Low Temp. Phys. {\bf 105}, 317 (1996).

\bibitem{koike4}
Y. Koike, S. Takeuchi, Y. Hama, H. Sato, T. Adachi, and M. Kato,
 Physica C {\bf 282-287}, 1233 (1997).

\bibitem{adachi2}
T. Adachi, S. Yairi, K. Takahashi, Y. Koike, I. Watanabe, and K. Nagamine,
 unpublished.

\bibitem{sato}
H. Sato, A. Tsukada, M. Naito, and A. Matsuda,
 Phys. Rev. B {\bf 62}, R799 (2000).
 
\bibitem{wells}
B. O. Wells, R. J. Birgeneau, F. C. Chou, Y. Endoh, D. C. Johnston, M. A. Kastner, Y. S. Lee, G. Shirane, T. M. Tranquada, and K. Yamada,
 Z. Phys. B {\bf 100}, 535 (1996).

\bibitem{wells2}
B. O. Wells, Y. S. Lee, M. A. Kastner, R. J. Christianson, R. J. Birgeneau, K. Yamada, Y. Endoh, and G. Shirane,
 Science {\bf 277}, 1067 (1997).

\bibitem{blakeslee}
P. Blakeslee, R. J. Birgeneau, F. C. Chou, R. Christianson, M. A. Kastner, Y. S. Lee, and B. O. Wells,
 Phys. Rev. B {\bf 57}, 13915 (1998).

\end{references}
\end{document}